\documentclass{PoS}
\usepackage{graphicx}% Include figure files
\usepackage{dcolumn}% Align table columns on decimal point
\usepackage{bm}% bold math
\usepackage{ulem}
\def  \bef  {\begin{figure}}
	\def  \eef  {\end{figure}}
\def  \be   {\begin{equation}}
\def  \ee   {\end{equation}}
\def  \ba   {\begin{array}}
	\def  \ea   {\end{array}}
\def  \bea  {\begin{eqnarray}}
\def  \eea  {\end{eqnarray}}
\def  \beq  {\begin{eqnarray}}
\def  \eeq  {\end{eqnarray}}
\def  \nn   {\nonumber}
\def  \bd   {\begin{displaymath}}
\def  \ed   {\end{displaymath}}
\def  \bse  {\begin{subequations}}
	\def  \ese  {\end{subequations}}
\def  \bwt  {\begin{widetext}}
	\def  \ewt  {\end{widetext}}

\def  \ba   {{\bf{a_1}}}

\title{Dynamical quenching weights in an expanding medium}

\ShortTitle{Dynamical quenching weights in an expanding medium}

\author{\speaker{Souvik Priyam Adhya}\\
        Institute of Particle and Nuclear Physics, Faculty of Mathematics and Physics, Charles University, V Holesovickach 2, Prague, 18000, Czech Republic\\
        E-mail: \email{souvik.priyam.adhya@cern.ch}}
\author{Carlos A. Salgado\\
        Departamento de Fisica de Particulas and IGFAE, Universidade de Santiago de Compostela, 15782 Santiago de Compostela, Spain\\}
\author{Konrad Tywoniuk\\
        Department of Physics and Technology, University of Bergen, Postboks 7803, 5020 Bergen, Norway\\}
\abstract{In this work, we extend the resummation of multiple medium-induced emissions to apply to dynamically expanding media. This is done by recasting the quenching weight as the solution of a rate equation with medium-induced partonic splitting functions that are sensitive to the expansion. We perform the calculations in the framework of Baier-Dokshitzer-Mueller-Peigne-Schiff-Zakharov (BDMPSZ) formalism for multiple soft scatterings with a time-dependent transport coefficient. Furthermore, we discuss the validity of a dynamical scaling law that relates the spectrum in an expanding medium to the equivalent static case with re-scaled medium parameters and test the scaling law for the gluon splitting rates.}

\FullConference{International Conference on Hard and Electromagnetic Probes of High-Energy Nuclear Collisions\\
		30 September - 5 October 2018\\
		Aix-Les-Bains, Savoie, France}

\begin{document}

\section{Introduction}
Medium induced radiative energy loss for highly energetic jet constituents has proven to be one of the most important mechanisms required to characterize hot and dense deconfined QCD matter. In fact, high $p_T$ hadronic observables are modified due to the interplay of soft and hard processes of energy loss at partonic level inside the medium. It has been established that a fast parton produced by hard process radiates gluons in course of interactions with a dynamically expanding medium, thus transfering significant energy at large angles due to multiple soft scattering within the dynamically expanding medium \cite{Salgado:2003gb,Wiedemann:2000za, Mehtar-Tani:2014yea, Arnold:2008iy, blaizot2013}. As a consequence, the jet spectrum and internal jet shape is modified leading to a renewed interest for the study of jet quenching parameters for evolving partonic matter for Monte-Carlo event generators.
In this work, we present results  confirming the validity of a dynamical scaling law, first derived in \cite{Salgado:2002cd} for the gluon spectrum and extending it to the gluon splitting rates for an expanding medium. We outline the methodology for the formulation of the medium induced splitting function through the kinematic rate equation to understand the quenching of jets for a dynamically extending partonic medium.
\section{Formalism}%%%%%%%%%%%%%%%%%%%%%%%%%%%
We start with the gluon emission spectra induced by multiple soft scattering in the partonic medium and is given as \cite{Wiedemann:2000za,Salgado:2003gb},
%
%\vspace{-0.5cm}
\begin{eqnarray}
&&\omega\frac{dI}{d\omega}
= {\alpha_s\,  C_R\over (2\pi)^2\, \omega^2}\,
2{\rm Re} \int_{\xi_0}^{\infty}\hspace{-0.3cm} dy_l
\int_{y_l}^{\infty} \hspace{-0.3cm} d\bar{y}_l\,
 \int d{\bf u}\,  \int_0^{\chi \omega}\, d{\bf k}_\perp\, 
e^{-i{\bf k}_\perp\cdot{\bf u}}   \,
e^{ -\frac{1}{2} \int_{\bar{y}_l}^{\infty} d\xi\, n (\xi)\sigma(\bf u)}\,
\nonumber \\
&& \times {\partial \over \partial {\bf y}}\cdot
{\partial \over \partial {\bf u}}\,\int {\cal D}{\bf r}
\exp\left[ i \frac{\omega}{2} \int_{y_1}^{y_2} d\xi 
\left(\dot{\bf r}^2- \frac{\Omega_\alpha^2(\xi_0)}{\xi^\alpha}\, {\bf r}^2 \right)\right]\, .
\label{2.1}
\end{eqnarray}
which is valid at small transverse distances $r=|r|$ and under the dipole approximation $\sigma(\bf{r})\propto\bf{r}^2$.
The path integral ${\cal K}({\bf r}_1,y_1;{\bf r}_2,y_2|\omega)$ (the last integral) is identical to the path integral of
a 2-dimensional harmonic oscillator with time-dependent imaginary
frequency.
In fact, the time dependence of the transport co-efficient $\hat{q}(\xi)$  and the equivalent static transport co-efficient $\bar{\hat{q}}$ can be expressed by
the following power law \cite{Salgado:2003gb},
\beq
\frac{\Omega_\alpha^2(\xi_0)}{\xi^\alpha} 
&=& \frac{\hat{q}(\xi)}{i\,2\,\omega}\\
%= -i \frac{\hat{q}_0}{2\omega} \left( \frac{\xi_0}{\xi}\right)^\alpha\, .
\hat{q}(\xi)=\hat{q_0}(\frac{\xi_0}{\xi})^\alpha;&&~~~~\bar{\hat{q}}=\frac{2}{L^2}\int_{\xi_0}^{L+\xi_0}d\xi(\xi-\xi_0)\hat{q}(\xi)
\label{c.2}
\eeq
where $\alpha=0$ characterizes the static medium, $\alpha=1$ is attributed 
for a one-dimensional boost invariant longitudinal expanding partonic medium.
The solution of the path integral can be written as \cite{Salgado:2003gb, Wiedemann:2000za},
\begin{eqnarray}
{\cal K}({\bf r}_1,y_1;{\bf r}_2,y_2|\omega) =
\frac{i\, \omega}{2\pi D(y_1,y_2)}\, \hspace{-0.25cm}\,\times \exp\left[-
\frac{-i \omega}{2\, D(y_1,y_2)}
\left(c_1 {\bf r}_1^2+ c_2 {\bf r}2^2 -
2 {\bf r}_1\cdot {\bf r}_2\right)\right]\, .
\label{c.10}
\end{eqnarray}
with \cite{Salgado:2003gb},
\begin{eqnarray}
\hspace{-0.7cm}&&D(\xi,\xi') =
\frac{2\nu}{(2\nu\Omega_\alpha(\xi_0))^{2\nu}}\, (z z')^\nu\hspace{-0.25cm}~~\times\left[I_\nu(z)K_\nu(z')-K_\nu(z)I_\nu(z')\right]\, ,
\label{c.15}\\
\hspace{-0.7cm}&&c_1 = z\, \left(\frac{z'}{z}\right)^\nu
\hspace{-0.25cm}\left[I_{\nu-1}(z)K_\nu(z')+K_{\nu-1}(z)I_\nu(z')\right]\, ,
\label{c.16}\\
\hspace{-0.7cm}&&c_2 = z'\, \left(\frac{z}{z'}\right)^\nu
\hspace{-0.25cm}\left[K_\nu(z)I_{\nu-1}(z') + I_\nu(z)K_{\nu-1}(z')\right]\, 
\label{c.17}
\end{eqnarray}
where, $\nu=(2-\alpha)^{-1}$ , $z \equiv z(\xi)= 2  \nu \Omega_\alpha(\xi_0)\xi^{(1/2\nu)}$ and corresponding $z' \equiv z(\xi')$.
For $\alpha=0$, we recover the known form of the BDMPS-Z gluon
distribution function in the limit $R\rightarrow \infty$ $(R = \chi \omega_c L$, $\chi\rightarrow \infty$; $R\rightarrow \infty)$
for $\omega < \omega_c$ (soft) case as\cite{Salgado:2003gb, Mehtar-Tani:2014yea, Baier:1996sk},
\begin{eqnarray}
   \lim_{R\to \infty}\, 
   \omega \frac{dI}{d\omega} ^{(soft)}\simeq 
           \frac{2\alpha_s C_R}{\pi} \sqrt{\frac{\omega_c}{2\, \omega}}
                  ;~~~~~( \omega < \omega_c\,) 
\end{eqnarray}
Now, starting from the Eq.(\ref{2.1}), we can compute the analogous gluon spectra for the expanding medium in the un-constrained kinematical case $R\rightarrow \infty$ as, 
\bea
&&I^{in-in}= -2\frac{\alpha_s C_F}{\pi} Re[\int_{\xi_0}^{L+\xi_0}\int_{y_l}^{L+\xi_0}\frac{1}{D(\bar{y_l},y_l)^2}-\frac{1}{(\bar{y_l}-y_l)^2}]dy_l d\bar{y_l}\nn\\
&&I^{in-out}= -2\frac{\alpha_s C_F}{\pi} Re[\int_{\xi_0}^{L+\xi_0}\frac{-1}{D(\bar{y_l},y_l)c_1(\bar{y_l},y_l)}-\frac{1}{(L+\xi_0-y_l)}]dy_l \nn\\
&&\lim_{R\to \infty}\, \omega \frac{dI}{d\omega}^{(med, expand)} \hspace{-0.25cm}= [I^{in-in}+I^{in-out}]
\label{2.10}
\eea
where the $[in,out]$ terminology is based upon the limits of the $y$ integrals \cite{Wiedemann:2000za}.
In addition, we have also extended the computation to the case of constrained kinematics ($R\neq \infty)$ starting with equation (\ref{2.1}).
%where $D(\bar{y_l},y_l)$ and $c_1(\bar{y_l},y_l)$ have the above mentioned forms.
The gluon splitting rate can be computed by taking the differential of the BDMPS-Z equation, Eq.(\ref{2.10}) as well as that of constrained kinematical case; with respect to the length of the medium. 
The results are presented in Figs.(\ref{fig1} and \ref{fig3}).  It is to be noted that we have used an equivalent static transport co-efficient $\bar{\hat{q}}$ in order to compare to the spectrum in a static medium. Thus, we re-confirm the validity of the scaling law as in \cite{Salgado:2003gb, Salgado:2002cd}.
\section{Results and discussions}
\begin{figure}[tbp]
	\centering
	\includegraphics[width=7.2cm,height=5.4cm]{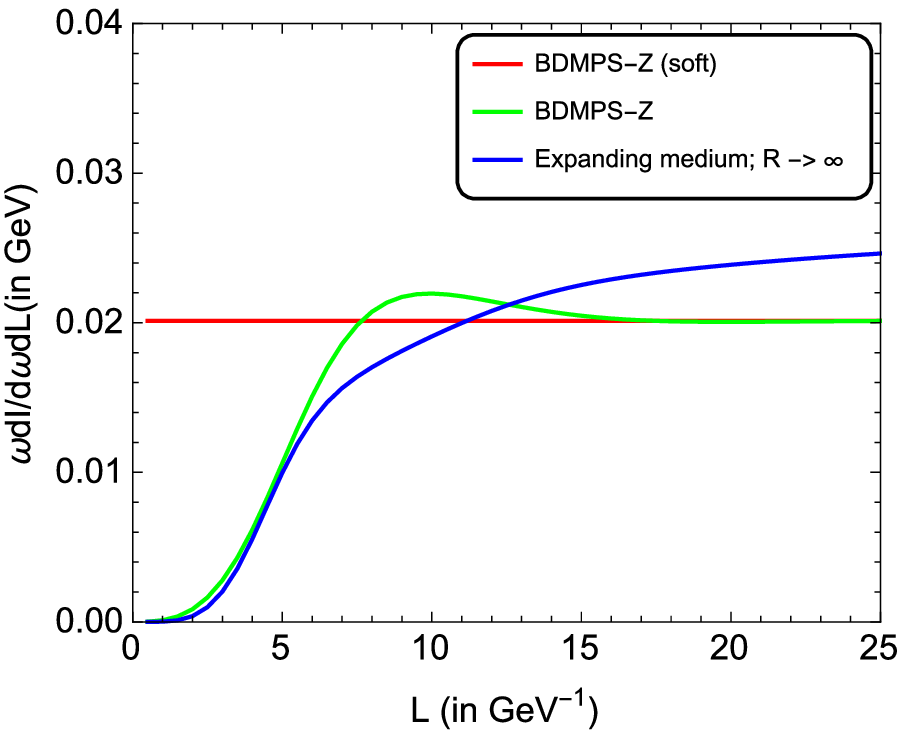}\includegraphics[width=7.2cm,height=5.4cm]{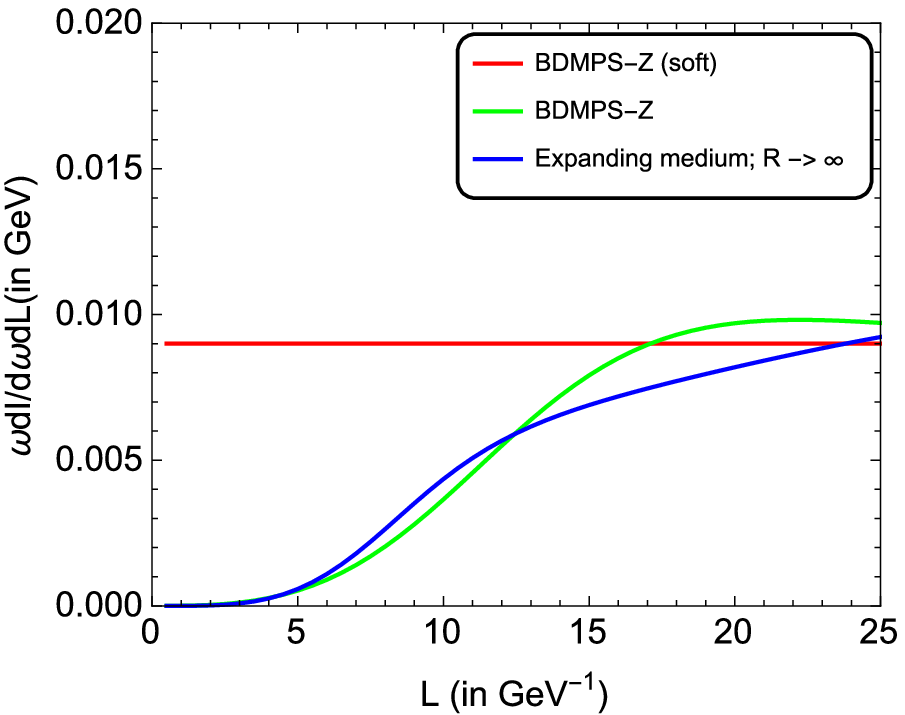}
	\caption{A comparison of the gluon splitting rate in a expanding medium with that of the BDMPS-Z rate for the unconstrained case for $\omega=2$ GeV (left) and $\omega=10$ GeV (right) respectively.The rate of the expanding medium has been produced with the equivalent static transport co-efficient.}
	\label{fig1}
\end{figure}

\begin{figure}[tbp]
	\centering
	\includegraphics[width=7.2cm,height=5.4cm]{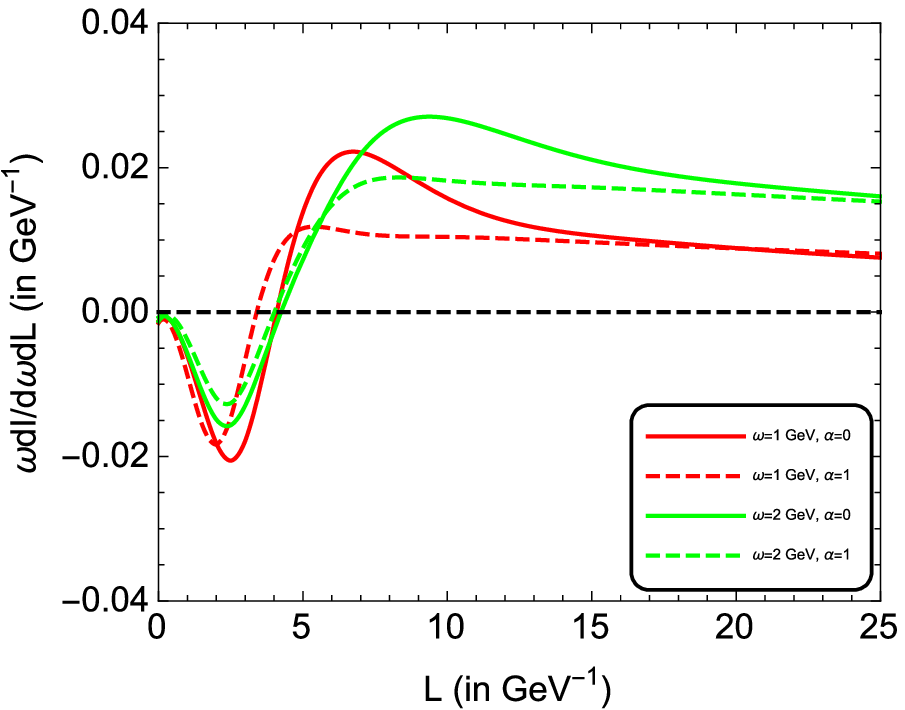}\includegraphics[width=7.2cm,height=5.4cm]{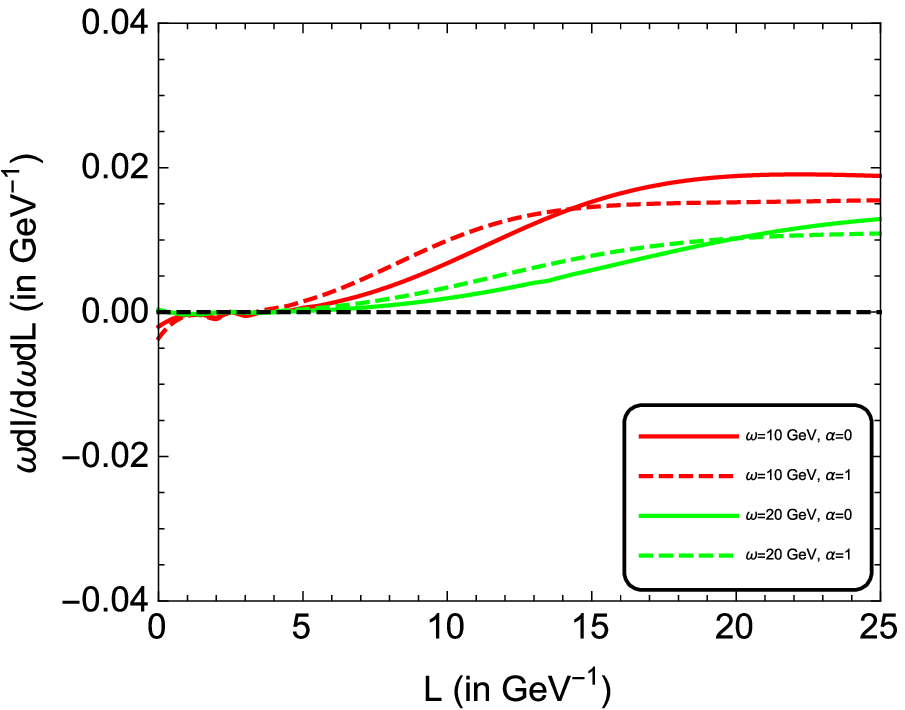}
	\caption{A comparison of the gluon splitting rate in a static medium with that of an expanding medium for the constrained case with scaling by the equivalent static transport co-efficient for $\omega=1, 2$ GeV (left) and $\omega=10, 20$ GeV (right) respectively .}
	\label{fig3}
\end{figure}

In Fig.(\ref{fig1}), we have presented the comparison of the gluon splitting rate derived from the BDMPS-Z spectra which is valid for the static medium with that of the expanding medium for the constrained kinematical scenario. For our purpose, we have explored the validity of the scaling law by using the equivalent static transport co-efficient for the expanding case $(\bar{\hat{q}}\sim\hat{q_0} L)$. The splitting rate for the case of small $\omega$ has also been plotted which is independent of the length of the medium. We see that there is a modest agreement between the two cases at values of $2$ and $10$ GeV in $\omega$ respectively.
In Fig.(\ref{fig3}), we have plotted the splitting rates for the static and expanding medium for the constrained kinematical case for different values of $\omega$. The validity of the scaling law is in good agreement as depicted in the plots.
%Fig.(\ref{fig3}) represents similar comparison as in Fig.(\ref{fig2}) for values of the gluon energy.
In this ongoing work, we have analysed the modification to the gluon spectra for the case of Bjorken expanding medium. We have presented results for the calculation of the gluon splitting rates for the static and expanding medium and discussed the validity of the scaling law for the equivalent transport co-efficient. Next, we plan to resum multiple medium induced radiation using the rate equation for expanding medium with defined kinematical constraints. Furthermore, we want to extend this work to study the medium modified intra-jet and out-of-jet distribution of the particles. In addition, we would also like to generalize to more complicated medium models like implementation of the hot spots in future. 

\textit{Acknowledgement:} K.T. is supported by the BFS Starting Grant. S.P.A. is  supported by Grant Agency of the Czech Republic under grant 18-12859Y and by Charles University grant UNCE/SCI/013. S.P.A. would like to thank M. Spousta for fruitful discussions.

\end{document}